\documentclass[11pt]{article}
\usepackage{graphicx}

\newcommand{\BABARPubYear}    {03}
\newcommand{\BABARConfNumber} {021}

\usepackage{rotating,amssymb}
\usepackage{rcdefs,subfigure}
\usepackage{axodraw}

\def\Mmiss{\ifmath{M_\mathrm{miss}}}
\def\Ecms{\ensuremath{E^\mathrm{CMS}}}
\def\pcms{\ensuremath{p^\mathrm{CMS}}}

\def\bdstardsstar {\ensuremath{\Bz\to\Dss\Dstarm}}
\def\nBB {{\ensuremath{N_{\BB}}}}
\def\dsphipi{\ensuremath{\Ds\rightarrow\phi\pip}}
\let\Dsphipi=\dsphipi

\input babarsym

\def\ee{\ifmath{e^{\sscr +}e^{\sscr-}}}
\def\ustat{\ensuremath{_\mathrm{stat}}}
\def\usyst{\ensuremath{_\mathrm{syst}}}
\def\mmax{\ensuremath{m_\mathrm{max}}}

\long\def\inst#1{\par\nobreak\kern 4pt\nobreak
    {\it #1}\par\vskip 10pt plus 3pt minus 3pt}

\begin{document}
{\pagestyle{empty}

\begin{flushright}
\babar-CONF-\BABARPubYear/\BABARConfNumber \\
August 2003 \\
\end{flushright}

\par\vskip 5cm

\begin{center}
\Large \bf Measurement of \boldmath$\BrFr(\Bz\to \Dss \Dstarm)$ and
Determination of the \boldmath\dsphipi\ Branching Fraction with
a Partial-Reconstruction Method
\end{center}
\bigskip

\begin{center}
\large The \babar\ Collaboration\\
\mbox{ }\\
\today
\end{center}
\bigskip \bigskip

\begin{center}
\large \bf Abstract
\end{center}
We present model-independent measurements of the branching fractions
${\cal B}(B^0\rightarrow D_s^{*+} D^{*-})$ and ${\cal B}(D_s^+ \rightarrow \phi \pi^+)$ based on 19.3 $\textup{fb}^{-1}$ of data
collected by the $BABAR$  detector at the PEP-II $e^+e^-$ $B$~Factory.
Neutral $B$-meson decays to the $D_s^{*+}D^{*-}$ final state are
selected with a partial reconstruction of the $D_s^{*+}$; that is, only the
$D^{*-}$ and the soft photon from the decay $D_s^{*+} \rightarrow \Ds \gamma$
are reconstructed. The branching fraction ${\cal B}(B^0\rightarrow D_s^{*+} D^{*-})$ is
extracted from these event yields, while ${\cal B}(D_s^+ \rightarrow \phi \pi^+)$ is determined by
combining this result with a previous measurement of the product
${\cal B}(B^0\rightarrow D_s^{*+} D^{*-}) \times {\cal B}(D_s^+ \rightarrow \phi \pi^+)$ with partial reconstruction
of the $D^{*-}$. We obtain the following preliminary results:
\begin{center}
  ${\cal B}(B^0\rightarrow D_s^{*+} D^{*-}) = (1.50 \pm 0.16 \pm 0.12)\%,$ \\
  ${\cal B}(D_s^+ \rightarrow \phi \pi^+)      = (4.7 \pm 0.6 \pm 0.8)\%$
\end{center}
where the first error is statistical, and the second systematic.

\vfill
\begin{center}
Contributed to the XXI$^{\rm st}$ International Symposium on Lepton and
Photon Interactions at High~Energies, 8/11 -- 8/16/2003, Fermilab, Illinois USA
\end{center}

\vspace{1.0cm}
\begin{center}
{\em Stanford Linear Accelerator Center, Stanford University, 
Stanford, CA 94309} \\ \vspace{0.1cm}\hrule\vspace{0.1cm}
Work supported in part by Department of Energy contract DE-AC03-76SF00515.
\end{center}

\newpage
} 

\begin{center}
\small

The \babar\ Collaboration,
\bigskip

%
B.~Aubert,
R.~Barate,
D.~Boutigny,
J.-M.~Gaillard,
A.~Hicheur,
Y.~Karyotakis,
J.~P.~Lees,
P.~Robbe,
V.~Tisserand,
A.~Zghiche
\inst{Laboratoire de Physique des Particules, F-74941 Annecy-le-Vieux, France }
% ...

\end{center}\newpage

\section{INTRODUCTION}
\label{sec:Introduction}
We present a measurement of the branching fractions
$\BrFr(\bdstardsstar)$ and $\BrFr(\Dsphipi)$ using a partial-reconstruction
technique\footnote[1]{Here and in the following,
charge-conjugate processes are implicitly considered.} \cite{CLEO}. 
A precise measurement of the branching fraction for
this mode is important because nearly all \Ds branching fractions are
determined by normalizing the measurements to $\BrFr(\Dsphipi)$
\cite{ref:pdg2002}. The present uncertainty on $\BrFr(\dsphipi)$ thus
affects many of the results regarding \Ds mesons, including the
determination of the decay constant by means of purely leptonic decays and
the measurement of the $\Ds\to K$ inclusive decay rate, as well as
$b$-physics analyses where a \Ds or a \Dss is fully reconstructed.

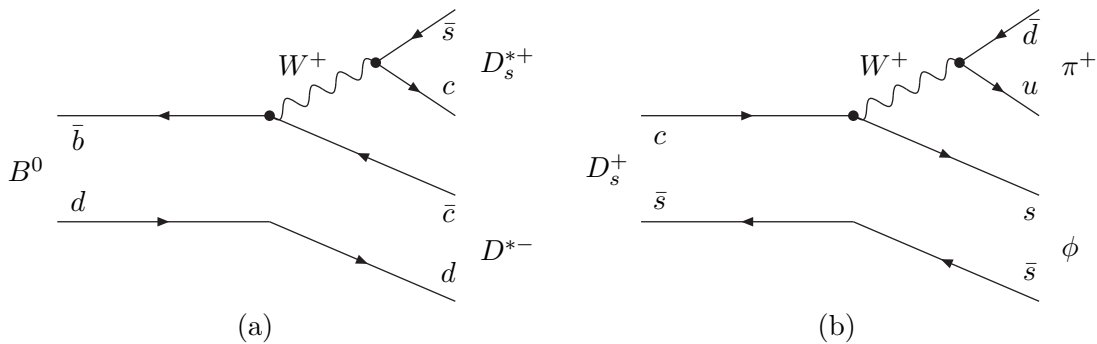
\begin{figure}[h!tbp] 
\begin{center}
\begin{picture}(405,150)(0,0)
\ArrowLine(10,40)(90,40)
\ArrowLine(90,40)(160,10)
\ArrowLine(90,80)(10,80)
\ArrowLine(160,50)(90,80)        \Vertex(90,80){2}
\Photon(130,100)(90,80){-3}{4}   \Vertex(130,100){2}
\ArrowLine(160,120)(130,100)
\ArrowLine(130,100)(160,80) 
\Text(18,48)[]{$d$}              \Text(158,91)[]{$c$}
\Text(158,21)[]{$d$}             \Text(18,72)[]{$\bar b$}
\Text(158,43)[]{$\bar c$}        \Text(158,112)[]{$\bar s$}
\Text(0,60)[]{\Bz}               \Text(103,100)[]{$W^+$}
\Text(170,100)[l]{\Dss}
\Text(170,30)[l]{\Dstarm}
\Text(85,0)[]{(a)}
\ArrowLine(310,40)(230,40)
\ArrowLine(380,10)(310,40)
\ArrowLine(230,80)(310,80)
\ArrowLine(310,80)(380,50)       \Vertex(310,80){2}
\Photon(350,100)(310,80){-3}{4}  \Vertex(350,100){2}
\ArrowLine(380,120)(350,100)
\ArrowLine(350,100)(380,80) 
\Text(238,48)[]{$\bar s$}        \Text(378,91)[]{$u$}
\Text(378,21)[]{$\bar s$}        \Text(238,72)[]{$c$}
\Text(378,43)[]{$s$}             \Text(378,112)[]{$\bar d$}
\Text(220,60)[]{\Ds}             \Text(323,100)[]{$W^+$}
\Text(390,100)[l]{\pip}
\Text(390,30)[l]{$\phi$}
\Text(305,0)[]{(b)}
\end{picture}
\end{center}
\caption{Tree-level Feynman diagrams for the decays (a) \bdstardsstar\
 and (b) \Dsphipi.}
\label{fig:feyn}
\end{figure}

In the factorization model for two-body decay rates, it is assumed that each
contribution to the transition amplitude of the process is
the product of two currents that can be evaluated separately.
This model has been successful~\cite{rosner} in describing the measured
branching fractions and polarizations for $B$ meson decays such as
$\Bz\to\Dstarm\pip$~\cite{dstarpai}, $\Bz\to\Dstarm\rho^+$ and
$\Bz\to\Dstarm a_1^+$~\cite{dstara1}, in which the momentum transfer in the 
process is low ($q^2\simeq M_\pi^2, M_\rho^2$).
Measurements of decay rates for modes such as \bdstardsstar\
(Fig.~\ref{fig:feyn}(a)) allow tests of the predictions made~\cite{luo} using 
the factorization model when the $W$ emits a light and a heavy quark and so
the momentum transfer is high ($q^2\simeq M_{D^*_s}^2$).

The Feynman diagram for the decay $\BrFr(\Dsphipi)$ is shown in
Fig.~\ref{fig:feyn}(b). 

\section{THE \babar\ DETECTOR AND DATASETS}
\label{sec:babar}
The data used in this analysis were collected with the \babar\ detector at
the \pep2\ storage ring and correspond to an
integrated luminosity of 19.3\ifb. A detailed description of the detector
can be found in Ref.~\cite{ref:babar}.

In addition to this data sample, several simulated event samples
were used in order to study efficiency and backgrounds.
For background studies, we used Monte Carlo samples of \BzBzb\
events (equivalent to an integrated luminosity of 270\ifb), \BBpm\
(150\ifb), \ee\to\ccbar (70\ifb) and \ee\to\uubar, \ddbar, \ssbar (70\ifb).
We used two signal samples in which the \bdstardsstar\ decay proceeds either
with completely longitudinal or transverse polarization; an additional
signal sample was extracted from the \BzBzb by selecting only
\bdstardsstar\ decays, with no further restriction on the \Dstarm and
\Dss decays.

\section{ANALYSIS METHOD}
\label{sec:Analysis}
\subsection{Analysis Strategy}
The partial reconstruction technique results in a significantly larger
sample of events than full reconstruction. Moreover, the
$\bdstardsstar\to(\Ds\gamma)(\Dzb\pim)$ decay is interesting from an
experimental point of view since it can be partially reconstructed in two
ways: the \Dss can be fully reconstructed and combined with the slow pion
from the decay $\Dstar\to\Dzb\pim$, or the \Dstarm can be fully
reconstructed and combined with the soft photon from the decay
$\Dss\to\Ds\gamma$.

The former technique has been used in \babar\,\cite{BAD524} to measure
$\BrFr(\bdstardsstar)$. However, the precision one can achieve through this
technique is limited by uncertainty on $\BrFr(\Dsphipi)$. By applying this
method, the \bdstardsstar\ branching fraction can be expressed as
\begin{equation}
\label{eq:brone}
  \BrFr(\bdstardsstar)={1\over 2 \nBB}
  {{N_{\Dss\pim}}\over{\BrFr(\Dstarm\to\Dzb\pim)
  \BrFr(\Dss\to\Ds\gamma) \sum_i (\eps_i \cdot \BrFr^{\Ds}_i)}},
\end{equation}
where $N_{\Dss\pim}$ is the number of partially reconstructed \Dstarm
candidates, \nBB\ is the number of neutral \B\ meson pairs,
$\BrFr^{\Ds}_i$ are the \Ds branching fractions, $\eps_i$ are the
total reconstruction efficiencies\footnote[2]{Both exclusive \Dss
reconstruction and partial reconstruction efficiencies are included in
$\eps_i$.} computed separately for each \Ds decay mode, and the index
$i$ runs over all \Ds decay modes used in the reconstruction (in
Ref.~\cite{BAD524}, \Dsphipi, $\Ds\to K^{*+}K^{0}$, and $\Ds\to
K^{*0}K^+$). Partial \Dss reconstruction similarly yields
\begin{equation}
\label{eq:brtwo}
  \BrFr(\bdstardsstar)={1\over 2\nBB}
  {{N_{\Dstarm\gamma}}\over{\BrFr(\Dss\to\Ds\gamma)
  \BrFr(\Dstarm\to\Dzb\pim)\sum_j (\eps_j \cdot \BrFr^{\Dz}_j)}},
\end{equation}
where $N_{\Dstarm\gamma}$ is the number of partially reconstructed \Dss
candidates. The result now depends on $\BrFr^{D^0}_j$, the branching
fractions for the \Dz modes, which are measured much more precisely than
those of the \Ds.

The \Dsphipi\ branching fraction can be extracted by combining
the two methods. Dividing Eq.\,\ref{eq:brone} by Eq.\,\ref{eq:brtwo} and
solving for $\BrFr(\Dsphipi)$, this last quantity can be expressed as
\begin{equation}
\label{eq:brdsphipi}
{{\BrFr(\Dsphipi)}=
{{N_{\Dss\pim}}\over{\sum_i{(\eps_i \cdot  R^{\Ds}_i)} }}\
{{\sum_j{(\eps_j \cdot \BrFr^{\Dz}_j)} }\over{N_{\Dstarm\gamma}}}},
\end{equation}
where $R^{\Ds}_i\equiv{\BrFr^{\Ds}_i/ \BrFr(\Dsphipi)}$ is the branching
fraction of each \Ds mode relative to the \Dsphipi\ mode, and the $\Dss\pim$
($\Dstarm\gamma$) yields are normalized to the same luminosity. In the
systematic uncertainty determination the contributions given by 
$\BrFr(\Dstarm\to\Dzb\pim)$ and $\BrFr(\Dss\to\Ds\gamma)$ are clearly
cancelled according to Eq.~\ref{eq:brdsphipi}.

\subsection{Signal Extraction}
We reconstruct the
$\bdstardsstar\to(\Ds\gamma)(\Dzb\pim)$ decay by combining photons in the
event with fully reconstructed \Dstarm mesons, without requiring
reconstruction of the \Ds from the \Dss decay. In order to extract the
signal, we compute the missing mass \Mmiss\ recoiling against the
\Dstarm-$\gamma$ system

\begin{equation}
\Mmiss = \sqrt{(E_\mathrm{beam} - E_{\Dstar} - E_{\gamma})^2 - 
(\overrightarrow{p}_B - \overrightarrow{p}_{\Dstar} - 
\overrightarrow{p}_{\gamma})^2}.
\label{eq:nostramm}
\end{equation}
For signal events, this must be the \Ds mass within experimental
resolution. The kinematics of the event are not fully constrained with this
procedure and one of the decay parameters must be chosen in an arbitrary
way. In particular, approximating the energy of the \B\ meson in the \ee\
center-of-mass (CM) to the CM beam energy, the opening angle between the
\B\ momentum vector and the measured \Dstar\ momentum vector can be
calculated from 4-momentum conservation in the \bdstardsstar\ decay
\begin{equation}
\cos\vartheta_{B\Dstar} = \frac{m^2_B-m^2_{\Dss}-2E_B E_{\Dstar}}{2|\vec
p_B||\vec p_{\Dstar}|}.
\end{equation}
The \B\ four-momentum is therefore determined up to the azimuthal angle
around the \Dstar\ direction. However, an arbitrary choice of the azimuthal
angle ($\cos\phi_{B\Dstar} = 0$) introduces only a negligible spread 
(less than 1.5 \MeVcc) in the missing mass distribution.

\subsection{Event Selection}
To reject events from continuum, we require the ratio of the
second to the zeroth Fox-Wolfram moment ($R_2$) \cite{R2} to be less than 0.3.

Candidates for \Dstarm are reconstructed in the $\Dzb\pim$ mode, using \Dzb
decays to $\Kp\pim$, $\Kp\pim\pipi$ $\Kp\pim\pio$, and
$\Kos\pipi$, here listed in order of decreasing purity. The $\chi^2$
probabilities of both the \Dz and \Dstar vertex fits are required
to be greater than 1\%. The \Dstar momentum in the \FourS frame
must satisfy $1.4\GeVc < \pcms(\Dstarm) < 1.8\GeVc$.
Moreover, we require the reconstructed mass of the \Dz particle 
to be within 2.5 standard deviations of the \Dz nominal mass,
and the \Dstarm Q-value ($Q(\Dstarm) \equiv M(\Dstarm)-M(\Dz)-M(\pim)$) 
to satisfy $Q_{\mathrm{lo}} < Q(\Dstarm) < Q_{\mathrm{hi}}$, where 
$Q_{\mathrm{lo}}=4.00$ to $5.25$ \MeVcc\ and $Q_{\mathrm{hi}}=6.75$ to 
$8.00$ \MeVcc, depending on the \Dz decay mode.
Kaon identification is required for the modes $\Kp\pim\pio$ and
$\Kp\pim\pipi$. For the mode $\Kos\pipi$, the invariant mass of the \pipi
from the \Kos\ decay is required to lie within 15\MeVcc\ of the \Kos\
nominal mass and its flight length must be greater than 3 mm.
If more than one \Dstar\ candidate is found, for each \Dz decay mode we
first select the candidates in which the pion from the decay
$\Dstarm\to\Dzb\pim$ has hits in the drift
chamber. Among these, the one with the minimum value of
$\chi^2 = 
  {{[(Q(\Dstarm) - Q_{\scr PDG}(\Dstarm))}/\sigma_{Q(\Dstarm)}}]^2
+ {{[(M(\Dz) - M_{\scr PDG}(\Dz))}/\sigma_{M(\Dz)}}]^2$ is
retained. Finally, if candidates from different \Dz decay modes are
present, we select the one with the best average purity.

The selection of photon candidates is based on the optimization
of the statistical significance of the observed signal
($S/\sqrt{S+B}$, where $S$ and $B$ are the
number of signal and background photons), using generic Monte Carlo events. 
We apply a \piz veto on photon candidates, rejecting them if their
invariant mass, calculated with any other photon candidate in the event,
is between 115 and 155\MeVcc.
The following additional cuts are applied on the photon energy in the
\FourS CMS (\Ecms), the cluster lateral moment (LAT) \cite{physbook} and
Zernike moments \cite{zernike} of order $\{2,0\}$ ($Z_{20}$) and $\{4,2\}$
($Z_{42}$): {$\Ecms > 142\MeV$}, {$0.016 < LAT <0.509$}, {$Z_{20}>0.85$},
{$Z_{42}<0.14$}.
If more than one photon is found in the event, we choose the one which
maximizes the value of a likelihood ratio based on four photon variables 
($E$, \Ecms, $N_{cry}$, LAT), where $E$ is the photon energy in the
laboratory frame and $N_{cry}$ is the number of calorimeter crystals
involved in the electromagnetic shower.

\subsection{Selection efficiency and Monte Carlo validation}
\label{sec:efficiency}
To effect the partial \Dss reconstruction in Monte Carlo events, 
the Monte Carlo sample is split in two parts. The signal reconstruction
efficiency is determined from \bdstardsstar\ events extracted from the
first half of the sample by performing a minimum-$\chi^2$ fit to the
missing mass distribution. The signal peak, centered on the nominal \Ds\
mass, is well described by a Gaussian probability density function
(p.d.f.), while the background, which is mainly due to random
$\Dstar-\gamma$ combinations, is parametrized with the function
$B(m)=a\left(1-e^{-b(m-\mmax)}\right)\left(\frac{m}{\mmax}\right)^c$,
where $m \equiv \Mmiss$ and \mmax\ is the end point of the missing mass
distribution. We perform a single fit to all \Dz decay modes; the sum of the
branching fraction-weighted efficiencies for the four reconstruction modes
is computed from the number of signal events fitted in the range
$|m-M(\Ds)| < 41 \MeVcc$, and found to be $\vev{\eps \BrFr} =
\sum_j{(\eps_j \cdot \BrFr^{\Dz}_j)} = (7.14\pm0.16)\times 10^{-3}$.

We have validated the fitting technique and the method of extracting the
signal on the other half of our Monte Carlo sample.
The distribution of the missing mass of partially reconstructed
\Bz candidates is shown in Fig.~\ref{fig:missmasstot} for \BzBzb (including
signal), \BBpm, and continuum Monte Carlo events.
The signal yield is extracted from a minimum-$\chi^2$ fit of the
missing mass distribution to a sum of the signal, described by a Gaussian
function, and the background, described by the p.d.f. introduced above.
From the signal yield, using Eq.~\ref{eq:brtwo}, we obtain
$\BrFr(\bdstardsstar)=(1.43\pm0.04)\%$, which is consistent with the value
of $1.41\%$ used in the generation of the Monte Carlo.
\begin{figure}[!htb]
\begin{center}
\includegraphics[height=8cm]{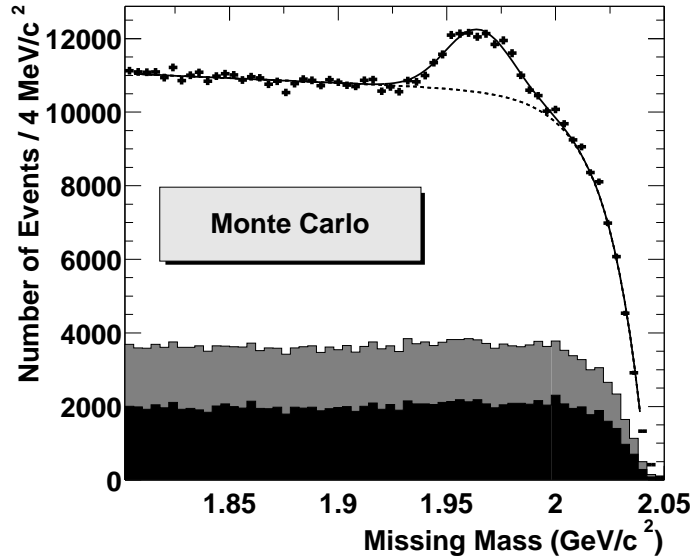}
\caption{Monte Carlo simulation of missing mass distributions. The missing
      mass is defined by Eq.~\ref{eq:nostramm}.
      Contributions of continuum (black), \BBpm\ (grey) and 
      \BzBzb\ (points) are added. The solid line shows the fit
      to the signal plus the sum of all backgrounds. The dashed line is
      the fit to the background component only.}
\label{fig:missmasstot}
\end{center}
\end{figure}

\subsection{Results on data}
Figure~\ref{fig:fitdata} shows the missing mass distribution in our data sample.
\begin{figure}[!htb]
\begin{center}
\includegraphics[height=8cm]{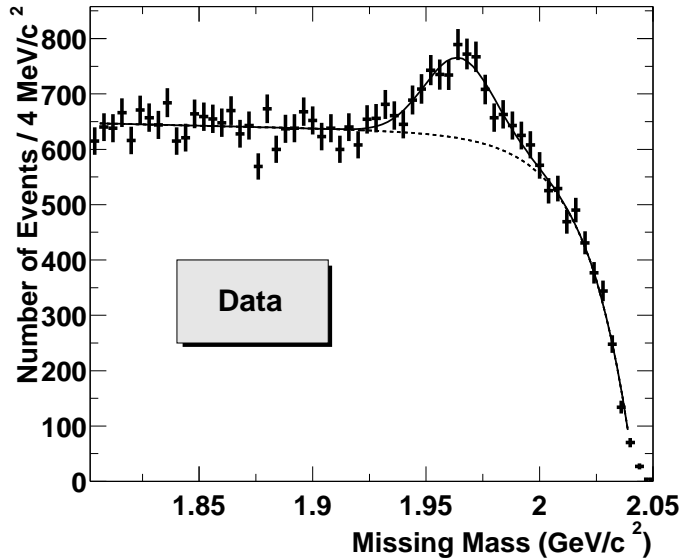}
\caption{Missing mass distribution in the data sample. The solid line shows
         the fit to the signal plus background, the dashed line is
         the fit to the background component only.}
\label{fig:fitdata}
\end{center}
\end{figure}
The same fitting procedure applied in the previous section to the Monte
Carlo sample is used to extract the number of signal events in the data
sample. In the fit we let all parameters float except the mean and the
stantard deviation of the Gaussian signal, which are fixed to their Monte Carlo values.
The result of the fit to the missing mass distribution is shown in
Fig.~\ref{fig:fitdata} as well. The signal yield in the data sample is
$N_{\Ds} = 1382 \pm 145$ events. The $\chi^2$ of the fit is 53.3 for 54
degrees of freedom, corresponding to a probability of 50.1\%.

From this yield we obtain $\BrFr(\bdstardsstar) = (1.50 \pm 0.16)\%$, where
the error is just statistical.

\section{SYSTEMATIC STUDIES}
\label{sec:Systematics}
The main sources of systematic uncertainties on the \bdstardsstar\ branching
fraction measurement are listed in Table~\ref{tab:syst}.
\begin{table}[htb]
\caption{Fractional systematic uncertainties (\%) for the \bdstardsstar\
branching fraction measurement.}
\begin{center}
\begin{tabular}{lr}
\hline
Source                       & Error $(\%)$ \\ \hline
Monte Carlo statistics       &       2.3    \\
Background shape             &       2.9    \\
$B$ counting                 &       1.1    \\ 
Tracking efficiency          &       2.4    \\ 
Soft pion efficiency         &       1.6    \\ 
Photon efficiency            &       4.2    \\
Particle identification      &       1.5    \\ 
Polarization uncertainty     &       0.8    \\ 
\Dz branching fractions      &       3.2    \\
$\BrFr(\Dstarm\to\Dzb\pim)$  &       0.7    \\
$\BrFr(\Dss\to\Ds\gamma)$    &       2.7    \\
\hline
Total systematic error       &       7.9   \\ \hline
\end{tabular}
\end{center}
\label{tab:syst}
\end{table}
The Monte-Carlo-statistics uncertainty is due to the statistical error
on the efficiency determination.
The uncertainty on the background shape is evaluated by fitting the
missing mass distribution using a different p.d.f. for the
background\footnote[3]{The alternative background p.d.f. has the following
functional form: $\displaystyle{B(m)=\frac{a(m-\mmax)^b}{c+(m-\mmax)^b}}$.}, 
and assigning the relative signal yield difference as systematics. The
systematic uncertainty due to tracking efficiency is evaluated applying a
random inefficiency of 0.8\% per track (1.6\% for the soft pions from
\Dstar decays). We assign as an uncertainty the difference between the yield
obtained in this way and the one described in Sec.\,\ref{sec:efficiency}.
The systematics associated to photon reconstruction efficiency
and particle identification are evaluated in a similar way.
We find a 7\% difference in the overall selection efficiency between our
samples with complete longitudinal or transverse polarization in the
\bdstardsstar\ decay. The uncertainty due to the dependence on polarization
is computed taking into account the measured value~\cite{BAD524} of the 
fraction of longitudinal
polarization and its uncertainty $\Gamma_L/\Gamma=(51.9\pm5.7)\%$.
Finally, the uncertainties on \Dz, \Dstarm and \Dss branching 
fractions~\cite{ref:pdg2002} are propagated through the analysis.

\section{PRELIMINARY BRANCHING FRACTION RESULTS}
\label{sec:Physics}
The \bdstardsstar\ branching fraction is found to be:
\begin{equation}
  \BrFr(\bdstardsstar) = (1.50 \pm 0.16 \pm 0.12)\%,
  \label{eq:ilnostro}
\end{equation}
where the first error is statistical, and the second systematic.
The \Dsphipi\ branching fraction can be extracted by comparing this result
with the measurement of the \bdstardsstar\ decay with partial \Dstarm
reconstruction \cite{BAD524}:
$\BrFr(\bdstardsstar) =(1.97\pm0.15\ustat\pm0.30\usyst\pm 0.49_{\dsphipi})\%$.
In this measurement the world average branching fraction
$\BrFr(\Dsphipi)=(3.6\pm0.9)\%$ was used.
From Eq.\,\ref{eq:ilnostro} we obtain therefore:
\begin{equation}
  \BrFr(\dsphipi)= (4.7 \pm 0.6 \pm 0.8)\%,
\label{eq:philast}
\end{equation}
where the first error is statistical, the second systematic.
The systematic uncertainty on this branching fraction is dominated by the
measurement using partial \Dstarm reconstruction.

\section{SUMMARY}
\label{sec:Summary}
A measurement of the \bdstardsstar\ branching fraction is performed,
using data corresponding to an integrated luminosity of 19.3\ifb,
with a partial reconstruction technique of $B$ meson:
$$
  \BrFr(\bdstardsstar) = (1.50 \pm 0.16 \pm 0.12)\%.
$$
This result is compatible with, and improves on the precision of previously
published experimental results~\cite{ref:pdg2002,BAD524}, and should be compared
with the most recent theoretical results based on the factorization
assumption~\cite{luo}: $\BrFr(\bdstardsstar)_\mathrm{theor} = (2.4 \pm 0.7)\%$.

The \Dsphipi\ branching fraction result is
$$
  \BrFr(\Dsphipi)= (4.7 \pm 0.6 \pm 0.8)\%.
$$
This new determination is compatible with the published CLEO
result~\cite{CLEO} and a preliminary measurement from Belle~\cite{Belle}.

\section{ACKNOWLEDGMENTS}
\label{sec:Acknowledgments}

We are grateful for the 
extraordinary contributions of our \pep2\ colleagues in
achieving the excellent luminosity and machine conditions
that have made this work possible.
The success of this project also relies critically on the 
expertise and dedication of the computing organizations that 
support \babar.
The collaborating institutions wish to thank 
SLAC for its support and the kind hospitality extended to them. 
This work is supported by the
US Department of Energy
and National Science Foundation, the
Natural Sciences and Engineering Research Council (Canada),
Institute of High Energy Physics (China), the
Commissariat \`a l'Energie Atomique and
Institut National de Physique Nucl\'eaire et de Physique des Particules
(France), the
Bundesministerium f\"ur Bildung und Forschung and
Deutsche Forschungsgemeinschaft
(Germany), the
Istituto Nazionale di Fisica Nucleare (Italy),
the Foundation for Fundamental Research on Matter (The Netherlands),
the Research Council of Norway, the
Ministry of Science and Technology of the Russian Federation, and the
Particle Physics and Astronomy Research Council (United Kingdom). 
Individuals have received support from 
the A. P. Sloan Foundation, 
the Research Corporation,
and the Alexander von Humboldt Foundation.


\begin{thebibliography}{99}

\bibitem{CLEO} The CLEO Collaboration, M. Artuso \etal, 
Phys.\ Lett. \textbf{B378}, 364 (1996).

\bibitem{ref:pdg2002}
Particle Data Group, 
K.~Hagiwara {\em et al.}, Phys.\ Rev.\ {\bf D66}, 010001 (2002).

\bibitem{rosner} 
J.~L.~Rosner, Phys.\ Rev.\ {\bf D42}, 3732 (1990).

\bibitem{dstarpai} 
The CLEO Collaboration, G.~Brandenburg \etal, 
Phys.\ Rev.\ Lett. {\bf 80}, 2762 (1998).

\bibitem{dstara1} 
The CLEO Collaboration, M.~S.~Alam \etal, Phys.\ Rev. {\bf D50}, 43 (1994).

\bibitem{luo} 
Z.~Luo and J.~L.~Rosner, Phys.\ Rev.\ {\bf D64}, 094001 (2001).

\bibitem{ref:babar}
The \babar\ Collaboration, B.\ Aubert {\em et al.},
Nucl.\ Instrum.\ Methods {\bf A479}, 1 (2002).

\bibitem{BAD524}
The \babar\ Collaboration, B.\ Aubert {\em et al.},
Phys.\ Rev.\ Lett. {\bf D67}, 092003 (2003).

\bibitem{R2} 
G.~C.~Fox and S.~Wolfram, Phys.\ Rev.\ Lett. {\bf 41}, 1581 (1978).

\bibitem{physbook}
P.~F.~Harrison and H.~R.~Quinn, {\it ``The \babar\ Physics Book''}, 
SLAC-R-504 (1998), p. 134.

\bibitem{zernike}
S.~P.~Prismall, M.~S.~Nixon and J.~N.~Carter, submitted to the 
{\it British Machine Vision Conference, Cardiff, United Kingdom, 2002},
74-82 (2002). 

\bibitem{Belle}
A.~Limosani, submitted to the {\it XXXVIII Rencontres de Moriond 
on Electroweak Interactions and Unified Theories, 
Les Arcs, France, 2003}, \texttt{hep-ex/0305037} (2003).
   
\end{thebibliography}
\end{document}